\documentclass[twocolumn,english,aps,pra,showpacs,superscriptaddress,floatfix]{revtex4-1}
\usepackage[latin9]{inputenc}
\usepackage{array}
\usepackage{units}
\usepackage{multirow}
\usepackage{amsmath}
\usepackage{amssymb}
\usepackage{graphicx}
\usepackage{esint}

\makeatletter

\providecommand{\tabularnewline}{\\}

\usepackage{babel}

\makeatother

\usepackage{babel}
\begin{document}

\title{Single-Nitrogen-vacancy-center quantum memory for a superconducting flux qubit mediated by a ferromagnet}

\author{Yen-Yu Lai}

\affiliation{Department of Physics and Center for Theoretical Physics, National
Taiwan University, Taipei 10617, Taiwan}

\affiliation{Center for Quantum Science and Engineering, National Taiwan University,
Taipei 10617, Taiwan}

\author{Guin-Dar Lin}

\affiliation{Department of Physics and Center for Theoretical Physics, National
Taiwan University, Taipei 10617, Taiwan}

\affiliation{Center for Quantum Science and Engineering, National Taiwan University,
Taipei 10617, Taiwan}

\author{Jason Twamley}

\affiliation{Centre for Engineered Quantum Systems, Department of Physics and
Astronomy, Macquarie University, NSW 2109, Australia}

\author{Hsi-Sheng Goan}
\email{goan@phys.ntu.edu.tw}

\affiliation{Department of Physics and Center for Theoretical Physics, National
Taiwan University, Taipei 10617, Taiwan}

\affiliation{Center for Quantum Science and Engineering, National Taiwan University,
Taipei 10617, Taiwan}

\date{\today}
\begin{abstract}
We propose a quantum memory scheme to transfer and store the quantum
state of a superconducting flux qubit (FQ) into the electron spin
of a single nitrogen-vacancy (NV) center in diamond via yttrium iron
garnet (YIG), a ferromagnet. Unlike an ensemble of NV centers, the
YIG moderator can enhance the effective FQ-NV-center coupling strength
without introducing additional appreciable decoherence. We derive
the effective interaction between the FQ and the NV center by tracing
out the degrees of freedom of the collective mode of the YIG spins.
We demonstrate the transfer, storage, and retrieval procedures, taking
into account the effects of spontaneous decay and pure dephasing.
Using realistic experimental parameters for the FQ, NV center and
YIG, we find that a combined transfer, storage, and retrieval fidelity
higher than 0.9, with a long storage time of 10 ms, can be achieved.
This hybrid system not only acts as a promising quantum memory, but
also provides an example of enhanced coupling between various systems
through collective degrees of freedom. 
\end{abstract}

\maketitle

\section{INTRODUCTION}

Superconducting qubits and related circuit-QED devices \cite{Clarke2008,Xiang2013}
with excellent scalability, parametric tunability, and strong coupling
with external fields are proving to be a powerful platform for quantum
information processing. However, they suffer from decoherence due
to inevitable interactions with their surrounding environments. In
a complex quantum protocol, superconducting qubits may experience
frequent idles times when they are not involved in active quantum
gates. During this idle time, to prevent the decoherence of their
quantum information, one can transfer their quantum state to an adjacent
quantum memory for better protection.

A hybrid system that takes advantage of the fast operation of superconducting
qubits and long coherence times of a suitable quantum memory may yield
good coherence preservation if the state transfer between them is
quick enough, i.e., faster than the decoherence time of either system.
The spin of a nitrogen-vacancy (NV) center in diamond, which has a
relatively long coherence time even at room temperature \cite{Doherty2013},
can be a candidate for such a quantum memory. This low decoherence
rate also means that the NV center normally only couples weakly to
a superconducting qubit. Such a weak coupling leads to a slow state
transfer and coherence loss can be significant. Ensembles of NV centers
\cite{Twamley2010,Kubo2010,Kubo2011,Lue2013,Zhu2011} may make the
coupling stronger, but at the added cost of increased decoherence
caused by internal spin-spin interactions, degrading the fidelity
of the quantum memory.

In this paper, we propose a scheme to transfer quantum states faithfully
between a superconducting flux qubit and a single-NV-center spin via
the ferromagnetic material yttrium iron garnet (YIG) \cite{Cherepanov1993,Serga2010}.
YIG has been proposed as a mediator for classical magnetic fields
to enhance the sensitivity of a NV magnetometer to achieve single
nuclear spin detection \cite{Trifunovic2015}. In addition, the large
number of spins in YIG with strong exchange interaction leads to collective-excitation
modes with narrow linewidths at low temperature \cite{Tabuchi2014,Zhang2015}.
These collective modes are known as quasiparticles or magnons \cite{Kruglyak2010},
and have been shown to be capable of coupling to different kinds of
quantum systems, such as superconducting microwave cavity modes \cite{Tabuchi2015,Zhang2015,Hisatomi2016}.
Magnons in YIG have also been proposed as a mediator of coherent coupling
between two distant spins (e.g., two spatially distant NV-center spins)
\cite{Trifunovic2013}. A CNOT gate between two single-NV-center spins
separated by a distance of about 1 $\mu$m with operation times of
the order of a few tens of nanoseconds has been demonstrated \cite{Trifunovic2013}.
This shows that a relatively strong coherent coupling between a single-NV-center
spin and YIG magnons is feasible. On the other hand, a flux qubit
(FQ) can display strong coherent coupling to an ensemble of NV centers
exhibiting a collective coupling of $\sim$70 MHz \cite{Zhu2011}.
However the spin density of YIG ($\rho\sim4.2\times10^{21}$ $\mathrm{cm}^{3}$)
\cite{Zhang2015} is almost three orders of magnitude larger than
typical NV ensembles ($\rho\sim5\times10^{18}$ $\mathrm{cm}^{3}$)
\cite{Zhu2011}. This suggests that the coupling between a flux qubit
and a small YIG sample may be similar or even stronger than between
a flux qubit and a NV ensemble. In this paper, we show that we can
achieve a substantially large effective coupling between a single-NV-center
spin and a FQ by using the magnons in a small nearby YIG sample as
a mediator without appreciably sacrificing the transfer and storage
fidelity of the quantum state.

When the size of the YIG is small enough, the Kittel mode (KM) of
the YIG sample \cite{Tabuchi2015,Zhang2015} is gapped from the higher-energy
modes and thus plays an important role in a low-temperature and low-excitation
regime. In our scheme, we find that the effective coupling and the
spatial separation between the FQ and NV required to attain these
coupling strengths via the YIG can be significantly enhanced. The
coupling attained using our proposal is of the order of several tenths
of MHz, while the spatial separation can be increased to a few tenths
of $\mathrm{\mu}$m. This represents an enhancement in the coupling
strength of 3\textendash 7 times over the direct FQ-NV coupling. More
interestingly, it also represents a substantial enhancement in spatial
separation required between the FQ and NV. For comparison, a direct
coupling scheme \cite{Douce2015} finds a coupling strength of $\sim$100
kHz, but requires a minuscule spatial separation of 20 nm. To achieve
larger direct coupling strength, strengths comparable to those found
using our scheme would require even tinier spatial separations, which
may be physically unrealistic. In contrast, in our proposal we are
able to expand the spatial separation to a few tenths of a $\mathrm{\mu}$m
scale, which is 10\textendash 20 times larger than the separation
required to attain similar coupling strengths via direct FQ-NV coupling.
Thus our scheme can provide significant couplings over a separation,
which is technically far easier to engineer. The quantum state transfer
time with the coupling strength found in our scheme is considerably
smaller than the decoherence time of the FQ so that fast and faithful
transfer can be realized without suffering significant decoherence.

The paper is arranged as follows. In Sec.~\ref{sec:II Model}, we
derive the effective Hamiltonian and coupling strength between a FQ
and a NV-center spin from a FQ-YIG-NV-center hybrid system. In Sec.~\ref{sec:III Quantum memory},
we introduce a protocol for the transfer and storage of the quantum
state. After that, simulations of the protocol are presented and discussed
in Sec.~\ref{sec:Results}, taking major decoherence effects into
consideration. Finally, a short conclusion is given in Sec.~\ref{sec:Conclusion}.
All the details of derivations of equations and calculations are presented
in Appendices~\ref{sec:Coupling with magnons} and \ref{sec:Derivation of the effective Hamiltonian by Schriffer-Wolff transformation}.

\section{MODEL}

\label{sec:II Model}

\begin{figure}
\includegraphics[width=1\columnwidth]{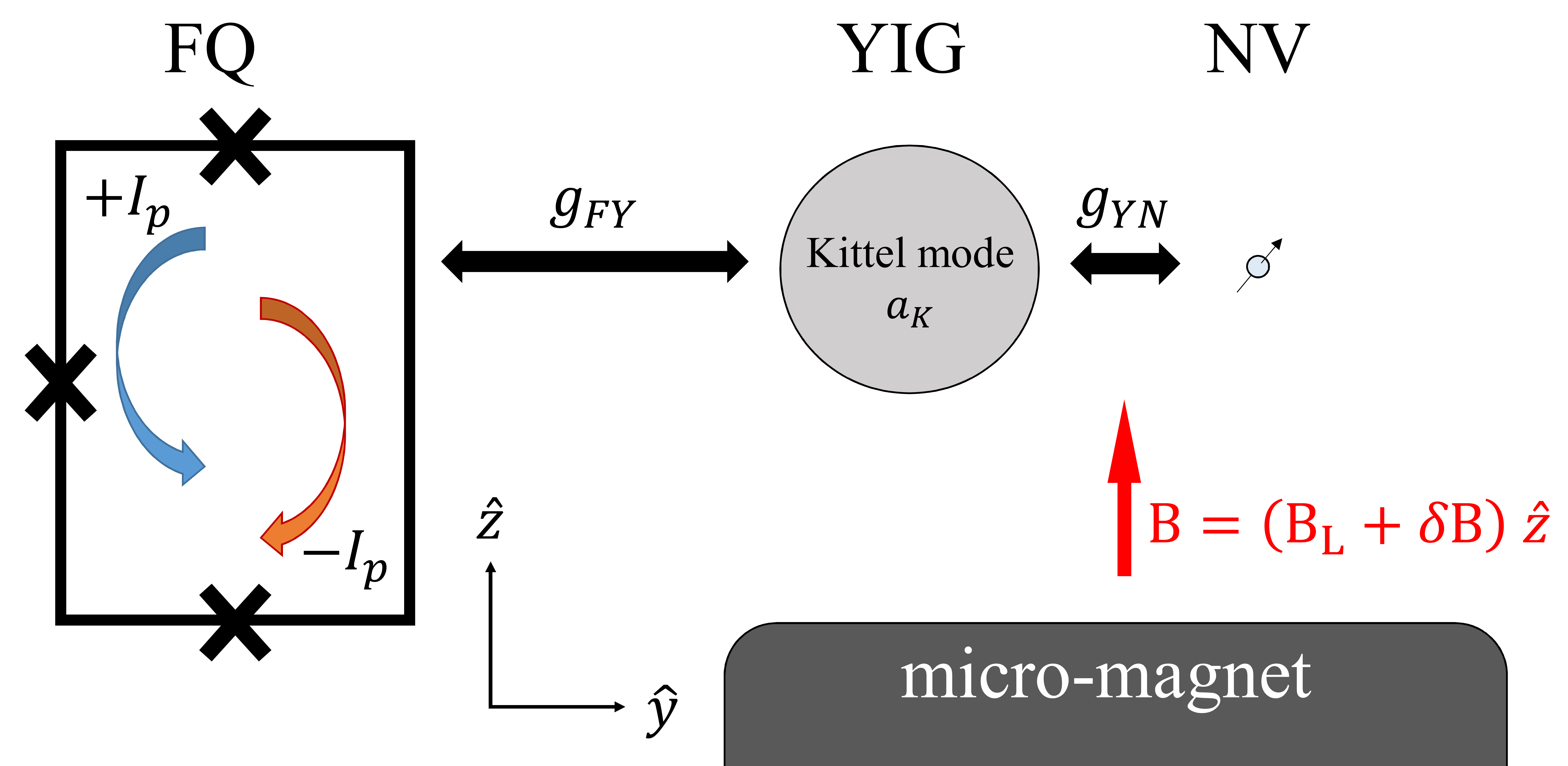} \caption{Schematic illustration of the proposed quantum memory. The FQ is regarded
as a two-level system depending on the sign (direction) of its persistent
current $I_{p}$, and the KM in YIG couples to both FQ and the single-NV-center
spin with strengths $g_{FY}$ and $g_{YN}$, respectively. The external
magnetic field felt by the YIG and the NV center is $B=B_{L}+\delta B$
(see Fig.~\ref{fig:2}), where $B_{L}$ is a local magnetic field
generated by a micromagnet without disturbing the FQ, and $\delta B$
is a tuneable magnetic field, generated by, e.g., a coil.}
\label{fig:1} 
\end{figure}

\begin{figure}
\includegraphics[width=1\columnwidth]{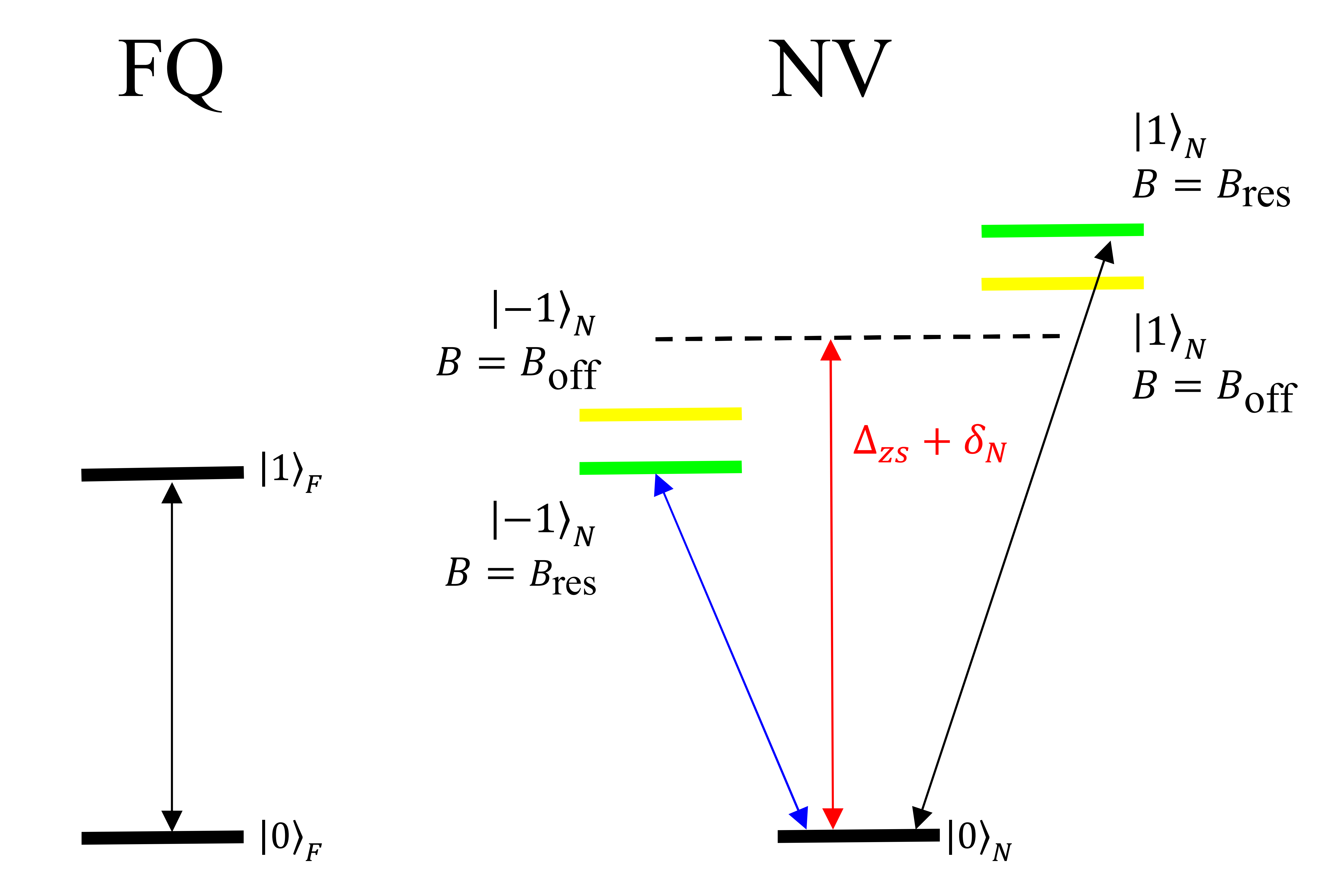} \caption{Energy-level diagram of the FQ (left) and the ground triplet states
of the NV-center spin (right). The NV-center spin states $\left|\pm1_{N}\right\rangle $
are degenerate (dashed line) and have a zero-field splitting $\Delta_{\mathrm{\mathrm{ZS}}}$
and an energy shift $\delta_{N}$ induced by the indirect coupling
scheme via YIG relative to state $\left|0_{N}\right\rangle $ (see
red arrow). By tuning the magnetic field on the NV-center spin to
the value of $B=B_{\mathrm{res}}$, one can control the transition
between $\left|0_{N}\right\rangle $ and $\left|-1_{N}\right\rangle $
to be resonant with the FQ (see blue arrow) or to be off-resonant
at $B=B_{\mathrm{off}}$. The transition between $\left|0_{N}\right\rangle $
and $\left|1_{N}\right\rangle $ is always set to be off-resonant
with the FQ, i.e., is set to be a disconnected channel. }
\label{fig:2} 
\end{figure}

The hybrid system in our proposal is schematically illustrated in
Fig.~\ref{fig:1} and contains three parts: the FQ, YIG, and a single-NV
center. The noninteracting Hamiltonian describing the individual systems
{[}with $\left(\hbar=1\right)${]} can be written as 
\begin{align}
H_{s}= & H_{F}+H_{Y}+H_{N},\label{eq:Hs}
\end{align}
where 
\begin{align}
H_{F}= & \frac{1}{2}\omega_{F}\sigma_{F}^{(z)},\label{eq:HF}\\
H_{Y}= & -J\sum_{\left\langle r,r'\right\rangle }S_{r}\cdot S_{r'}+\gamma_{e}B\sum_{r}S_{r}^{(z)},\label{eq:HY}\\
H_{N}= & \Delta_{\mathrm{ZS}}\left(S_{N}^{(z)}\right)^{2}-\gamma_{e}BS_{N}^{(z)}.\label{eq:HN}
\end{align}
Here, the FQ is regarded as a typical two-level system described by
the Hamiltonian $H_{F}$ in Eq.~(\ref{eq:HF}), with $\omega_{F}$
the transition frequency of the FQ and $\sigma^{(z)}$ the Pauli matrix
(for details, see Appendix \ref{sec:Coupling with magnons}). The
Hamiltonian of the YIG in an external field along the $z$ axis is
given by $H_{Y}$ in Eq.~(\ref{eq:HY}), where $S_{r}$ are the operators
of the spin located at position $r$ in the YIG. The parameter $J$
is the exchange coupling between the spins inside the YIG. We consider
the application of $B=B_{L}+\delta B$, an external magnetic field
along the $z$ axis and which is felt by the YIG and the NV center
(see Fig.~\ref{fig:1} and~\ref{fig:2}). Here, $B_{L}$ is a local
magnetic field generated by a micromagnet \cite{Norpoth2008} without
disturbing the FQ (for details, see Sec.~\ref{sec:Results}), and
$\delta B$ is the tuneable magnetic field whose value is set below
the critical field of the FQ. The tuneable dc magnetic field could
be generated by a coil. The ground triplet states of the NV center
is described by the Hamiltonian $H_{N}$ in Eq.~(\ref{eq:HN}), where
$S_{N}^{(z)}$ is the $z$ component of the spin-1 operators of the
single-NV center, $\Delta_{\mathrm{ZS}}=2.87$ GHz is the zero-field
splitting of the ground triplets, and $\gamma_{e}=-1.76\times10^{11}$
$\mathrm{rad}$ $\mathrm{s^{-1}}$ $\mathrm{T^{-1}}$ is the gyromagnetic
ratio of electron spin. To proceed further, let us simplify the Hamiltonians
a little bit. There are two single-photon transitions between $\left|0_{N}\right\rangle $
and $\left|\pm1_{N}\right\rangle $ in the ground triplet states of
the NV-center spin, and their energy gaps are $\omega_{(\pm1)}=\Delta_{\mathrm{ZS}}\mp\gamma_{e}B$.
We use, for instance, the NV-center spin states $\left|0_{N}\right\rangle $
and $\left|-1_{N}\right\rangle $ as our storage qubit basis state.
In the state transfer stage, this transition frequency is tuned to
be resonant with the FQ transition frequency, while the $\left|0_{N}\right\rangle $
to $\left|1_{N}\right\rangle $ transition is largely detuned (see
Fig.~\ref{fig:2}). Further, as shown later, the effective coupling
between the FQ and the NV-center storage qubit can be switched on
and off (or very small) by varying the external $B$ field. Here,
for the sake of deriving the effective Hamiltonian, we first treat
the NV-center spin as a two-level storage qubit by ignoring the far-detuned
transition. We will take into account the effect of the existence
of the far-detuned NV $\left|1_{N}\right\rangle $ state when we run
numerical simulations for the dynamics of quantum state transfer and
storage processes. We then transform the YIG Hamiltonian from a Heisenberg
model to a magnon form with the Holstein-Primakoff transformation
\cite{Holstein1940,Trifunovic2013,Tabuchi2016} and harmonic approximation.
As a result, the NV and YIG Hamiltonians can be rewritten as \cite{Trifunovic2013,Tabuchi2016}
\begin{align}
H_{N}\simeq & \frac{1}{2}\omega_{N}\sigma_{N}^{(z)},\\
H_{Y}\simeq & \sum_{k}\omega_{k}a_{k}^{\dagger}a_{k},\label{eq:Hs_rewrite}
\end{align}
where $\omega_{N}=\omega_{(-1)}$, and $\omega_{k}=sJa^{2}k^{2}+\gamma_{e}B$
are the frequencies of the NV storage qubit and magnon mode $k$,
respectively, $a_{k}^{\dagger}\left(a_{k}\right)$ is the creation
(annihilation) operator of magnon mode $k$, $s$ is the maximum eigenvalue
of the spin operator $S_{r}^{(z)}$ , and $a$ is the lattice constant
of the YIG. For a small-sized YIG sample, the boundary conditions
at the surface are of great importance and the magnon modes become
gapped.

The FQ and NV interact indirectly via YIG. The coupling Hamiltonian
thus has two parts: the FQ-YIG coupling and the YIG-NV coupling. The
current in the loop of the FQ generates a magnetic field which interacts
with the spins in the YIG. The NV electron spin also couples to these
spins by dipole-dipole interaction. Under the rotating-wave approximation
(RWA), the coupling Hamiltonian in terms of the YIG collective-excitation
modes (derived in detail in Appendix \ref{sec:Coupling with magnons})
reads 
\begin{eqnarray}
H_{c} & = & H_{FY}+H_{YN},\label{eq:Hc}\\
H_{FY} & \simeq & -\sum_{k}\left(g_{FY}(k)\sigma_{F}^{(+)}a_{k}+H.C.\right),\\
H_{YN} & \simeq & -\sum_{k}\left(g_{YN}(k)a_{k}^{\dagger}\sigma_{N}^{(-)}+H.C.\right),
\end{eqnarray}
where 
\begin{align}
g_{FY}(k)= & \frac{\mu_{0}}{2\pi}\gamma_{e}I_{p}\sqrt{\frac{2s}{N}}\sum_{r_{F}}^{N}\frac{e^{-ik\cdot a}}{r_{F}},\\
g_{YN}(k)= & -\frac{\mu_{0}}{4\pi}\gamma_{e}^{2}\hbar\sqrt{\frac{2s}{N}}\sum_{r_{N}}^{N}\left(\frac{3\mathrm{cos}^{2}\theta_{r_{N}}-1}{r_{N}^{3}}\right)e^{ik\cdot a},\label{eq:g0}
\end{align}
are the coupling strength of the FQ and NV with magnon mode $k$ in
the YIG, respectively. Here, $\mu_{0}$ is the vacuum permeability,
$I_{p}$ is the persistent current of the FQ, $r_{F}$ ($r_{N}$)
is the distance between a spin in the YIG and the FQ (the NV spin),
$N$ is the number of the spins in the YIG, and $\theta_{r_{N}}$
is the angle between the vector connecting the NV and the spin in
YIG, and the direction of the external magnetic field. Under the condition
that the Heisenberg interaction inside the YIG is much greater than
the coupling between a qubit and any single spin in the YIG, the qubit
then effectively interacts with the collective mode of all the spins
in the YIG. The more spins in the YIG following this condition, the
stronger the coupling.

If one chooses a YIG sphere with a submicrometer diameter, then the
energy levels of the YIG magnon modes are gapped, largely due to its
small size. We consider only the simplest mode of the YIG, i.e., the
KM \cite{Tabuchi2015,Zhang2015}, with frequency $\omega_{K}$ which
is far from both the frequencies of the FQ and NV-center storage qubit.
Thus the KM is in a virtual coupling regime with the FQ and NV-center
storage qubit. To account for the overall effect, we use the Schrieffer-Wolff
transformation (SWT) to derive the effective Hamiltonian up to the
second order in the coupling strengths with the YIG by averaging out
the far-off-resonance degrees of freedoms of the YIG. We then obtain
(with detailed derivation shown in Appendix \ref{sec:Derivation of the effective Hamiltonian by Schriffer-Wolff transformation})
an effective Hamiltonian between the FQ and the NV-center qubit as
\begin{align}
H_{\mathrm{eff}}\simeq & \frac{1}{2}\omega_{F,\mathrm{eff}}\sigma_{F}^{(z)}+\frac{1}{2}\omega_{N,\mathrm{eff}}\sigma_{N}^{(z)}\nonumber \\
 & +g_{FN,\mathrm{eff}}\left(\sigma_{F}^{(+)}\sigma_{N}^{(-)}+H.C.\right),\label{eq:H_eff}
\end{align}
where 
\begin{align}
\omega_{F,\mathrm{eff}}= & \omega_{F}+\delta_{F},\\
\omega_{N,\mathrm{eff}}= & \omega_{N}+\delta_{N},
\end{align}
are the effective frequencies of the FQ and the NV-center storage
qubit with frequency shifts $\delta_{F}=\frac{g_{FY}^{2}(\omega_{K})}{\omega_{F}-\omega_{K}}$
and $\delta_{N}=\frac{g_{YN}^{2}(\omega_{K})}{\left(\omega_{N}+\delta_{YN}\right)-\omega_{K}}$
induced by the KM of the YIG, respectively, and 
\begin{align}
g_{FN,\mathrm{eff}}(\omega_{K})= & \frac{1}{2}g_{FY}(\omega_{K})g_{YN}(\omega_{K})\nonumber \\
 & \times\left[\frac{1}{\omega_{F}-\omega_{K}}+\frac{1}{\left(\omega_{N}+\delta_{YN}\right)-\omega_{K}}\right]\label{eq:g_eff}
\end{align}
is the effective coupling strength between the FQ and the NV-center
qubit. Here, $\omega_{K}$ denotes the frequency of the KM. Note that
to have a substantially large effective coupling, the detuning $(\omega_{F}-\omega_{K})$
and detuning $(\omega_{N}+\delta_{YN}-\omega_{K})$ appearing in the
denominator of Eq.~(\ref{eq:g_eff}) should not be too large. On
the other hand, to keep the KM in a regime of virtual coupling with
the FQ and NV-center storage qubit, they should not be too small.
By varying the external magnetic field, we can control the values
of $(\omega_{F}-\omega_{K})$ and $(\omega_{N}-\omega_{K})$. In particular,
the change in $\omega_{K}$, due to the variation of the magnetic
field, is opposite to the change of the energy difference between
$\left|0_{N}\right\rangle $ and $\left|-1_{N}\right\rangle $ {[}in
contrast to the same energy change between $\left|0_{N}\right\rangle $
and $\left|1_{N}\right\rangle $ resulting in no change in $(\omega_{(+1)}-\omega_{K})${]}.
This makes $\left|0_{N}\right\rangle $ and $\left|-1_{N}\right\rangle $
a better choice of storage qubit basis states.

\section{QUANTUM MEMORY }

\label{sec:III Quantum memory}

Next we will use the derived effective Hamiltonian to investigate
the dynamics and the fidelity of the proposed quantum memory scheme.
There are two stages that we need to consider: state transfer stage
and state storage stage.

To better assess and calculate the fidelity of our scheme, we take
all the lowest triplet states of the NV spin into account. In this
case, the FQ couples to two transitions in these triplets separately
and the effective Hamiltonian, given by Eq.~(\ref{eq:H_eff}), becomes
\begin{align}
H_{\mathrm{eff}}= & \frac{1}{2}\omega_{F,\mathrm{eff}}\sigma_{F}^{(z)}+\sum_{j=\pm1}\omega_{N,(j),\mathrm{eff}}(B)\left|j_{N}\right\rangle \left\langle j_{N}\right|\nonumber \\
 & +g_{(+1)}\left[\sigma_{F}^{(+)}S_{N,(+1)}^{(-)}+H.C.\right]\nonumber \\
 & +g_{(-1)}\left[\sigma_{F}^{(+)}S_{N,(-1)}^{(-)}+H.C.\right],\label{eq:H_eff_3_level}
\end{align}
Here in the spin-1 Hilbert space of the NV center, the effective NV
spin frequencies are $\omega_{N,(\pm1),\mathrm{eff}}(B)=\Delta_{\mathrm{ZS}}\mp\gamma_{e}B+\delta_{N,(\pm1)}$,
where $\Delta_{\mathrm{ZS}}$ is the zero-field splitting and $\delta_{N,(\pm1)}$
is the frequency shift. The effective coupling strengths between the
FQ and the NV spin transitions are denoted as $g_{(\pm1)}$, corresponding
to Eq.~(\ref{eq:g_eff}) with $\omega_{N}\to\omega_{N,(\pm1),\mathrm{eff}}(B)$.
The subscripts $(\pm1)$ in the expression (and in the following),
stand for the transitions between $\left|0_{N}\right\rangle $ and
$\left|\pm1_{N}\right\rangle $, respectively. The operators $S_{N,(\pm1)}^{(\pm)}$
with superscript $\pm$ denote the raising and lowering operators,
respectively.

Now we move to the interaction picture through the unitary transformation
$U=\exp(-itH_{0,\mathrm{eff}})$, where $H_{0,\mathrm{eff}}=\frac{1}{2}\omega_{F,\mathrm{eff}}\sigma_{F}^{(z)}+\sum_{j=\pm1}\omega_{N,(j),\mathrm{eff}}(B_{\mathrm{res}})\left|j_{N}\right\rangle \left\langle j_{N}\right|$
is the first two terms of the effective Hamiltonian given by Eq.~(\ref{eq:H_eff_3_level})
with magnetic field $B=B_{\mathrm{res}}$, where $B_{\mathrm{res}}$
is the magnetic field strength applied to the NV-center spin when
the transition between $\left|0_{N}\right\rangle $ and $\left|-1_{N}\right\rangle $
matches the energy gap of the FQ (see Fig.~\ref{fig:2}). Then the
effective interaction Hamiltonian $H_{\mathrm{int}}$ becomes 
\begin{align}
H_{\mathrm{int}}= & H_{N,\mathrm{int}}+H_{FN,\mathrm{int}},\label{eq:H_int}\\
H_{N,\mathrm{int}}= & \sum_{j=\pm1}\delta_{B,(j)}\left|j_{N}\right\rangle \left\langle j_{N}\right|,\\
H_{FN,\mathrm{int}}= & g_{(-1)}\left[\sigma_{F}^{(+)}S_{N,(-1)}^{(-)}+H.C.\right],\nonumber \\
 & +g_{(+1)}\left[\sigma_{F}^{(+)}S_{N,(+1)}^{(-)}e^{2it\gamma_{e}B_{\mathrm{res}}}+H.C.\right],\label{eq:H_int_details}
\end{align}
and 
\begin{align}
\delta_{B,(\pm1)}= & \mp\gamma_{e}\left(B-B_{\mathrm{res}}\right).
\end{align}
When $\delta_{B,-1}$ is tuned to zero, i.e., $B=B_{\mathrm{res}}$,
the $g_{(-1)}$ coupling terms start to transfer the quantum state
from the FQ to the NV-center spin and the fast oscillating components
in the $g_{(+1)}$ terms can be effectively neglected.

We initially prepare the NV in the ground state, $\left|\psi_{N}(0)\right\rangle =\left|0_{N}\right\rangle $.
Suppose that the FQ is in a general state characterized by angles
$\theta$ and $\phi$. Then the joint state is 
\begin{align}
\left|\psi(0)\right\rangle = & \left(\cos\theta\left|1_{F}\right\rangle +e^{i\phi}\sin\theta\left|0_{F}\right\rangle \right)\otimes\left|0_{N}\right\rangle \nonumber \\
= & \cos\theta\left|1_{F},0_{N}\right\rangle +e^{i\phi}\sin\theta\left|0_{F},0_{N}\right\rangle .\label{eq:state0}
\end{align}
After a transfer time $t=\pi/(2g_{(-1)})$, the target state in the
interaction picture becomes 
\begin{align}
\left|\psi(t)\right\rangle = & -i\cos\theta\left|0_{F},-1_{N}\right\rangle +e^{i\phi}\sin\theta\left|0_{F},0_{N}\right\rangle \nonumber \\
= & \left|0_{F}\right\rangle \otimes\left(-i\cos\theta\left|-1_{N}\right\rangle +e^{i\phi}\sin\theta\left|0_{N}\right\rangle \right).\label{eq:state1}
\end{align}

Once the state has been transferred to the NV-center spin, we turn
off the coupling effectively by enlarging the mismatch of the frequencies
between the FQ and the NV-center storage qubit. The quantum state
can thus be stored for better coherence with dephasing time characterized
by the NV-center spin's $T_{2}$ time. To retrieve the state from
the NV-center storage qubit to the FQ, we tune to the NV-FQ resonance
again. After a time $t_{1}=\pi/(2g_{(-1)})$, the original state is
restored in the FQ degrees of freedom 
\begin{equation}
\left|\psi(t_{f})\right\rangle =-e^{-i\phi_{s}}\cos\theta\left|1_{F},0_{N}\right\rangle +e^{i\phi}\sin\theta\left|0_{F},0_{N}\right\rangle ,\label{eq:state_final}
\end{equation}
with an additional phase $\phi_{s}$ that comes from the coherent
evolution during the storage time $t_{2}$ and since this is known
it can be corrected.

\section{Results and discussion}

\label{sec:Results}

We present the numerical results together with discussions to verify
our scheme here. Before proceeding with our numerical calculations
for the fidelity performance, we first describe the system parameters
used. It is assumed that the YIG is a sphere of radius about 45 $\textrm{nm}$
and contains about $10^{6}$ spins. A local magnetic field $B_{L}$
is generated by placing a micromagnet \cite{Norpoth2008} of size
$0.2\times0.2\times0.2$ $\mathrm{\mu m^{3}}$ with a uniform perpendicular
magnetization of about a hundred Gauss at a vertical distance $\sim$25\textendash 50
nm from the YIG. With this local magnetic field, the frequency of
the KM can reach GHz levels; furthermore, by varying an external magnetic
field the frequency difference $\left(\omega_{F}-\omega_{K}\right)$
can achieve a typical value of about 170 MHz. Since the direction
of $B_{L}$ is parallel to the plane of the FQ, the FQ is insensitive
to $B_{L}$. Furthermore, because the transverse distance from the
micromagnet boundary at which $B_{L}$ drops to 0, is smaller than
0.1 $\mathrm{\mu m}$ \cite{Norpoth2008}, if $r_{F}$ is considerably
larger than the sum of this transverse distance and the distance from
the YIG to the left boundary of the micromagnet (see Fig.~\ref{fig:1}),
we can realize local magnetic field control for the YIG without disturbing
the FQ. Therefore, by choosing a relatively large value of $r_{F}\sim$
0.25 $\mathrm{\mu}$m and using a FQ with $I_{p}=$500 nA and a diamond
with a single NV spin at a distance $r_{N}\sim$ 60 nm from the YIG,
we can estimate the effective coupling strength $g_{FN,\mathrm{eff}}$
to be about 350\textendash 700 kHz according to Eqs.~(\ref{eq:g_FY_details}),
(\ref{eq:g_YN_details}), and (\ref{eq:g_FN_eff_details}). The frequency
shifts of both the FQ and NV-center spin due to the YIG coupling {[}see
Eqs.~(\ref{eq:delta_F_details}) and (\ref{eq:delta_N_details}){]}
are at about hundreds kHz, and are thus rather small and negligible
in comparison with their own frequencies in the GHz range.

\begin{figure}
\includegraphics[width=1\columnwidth]{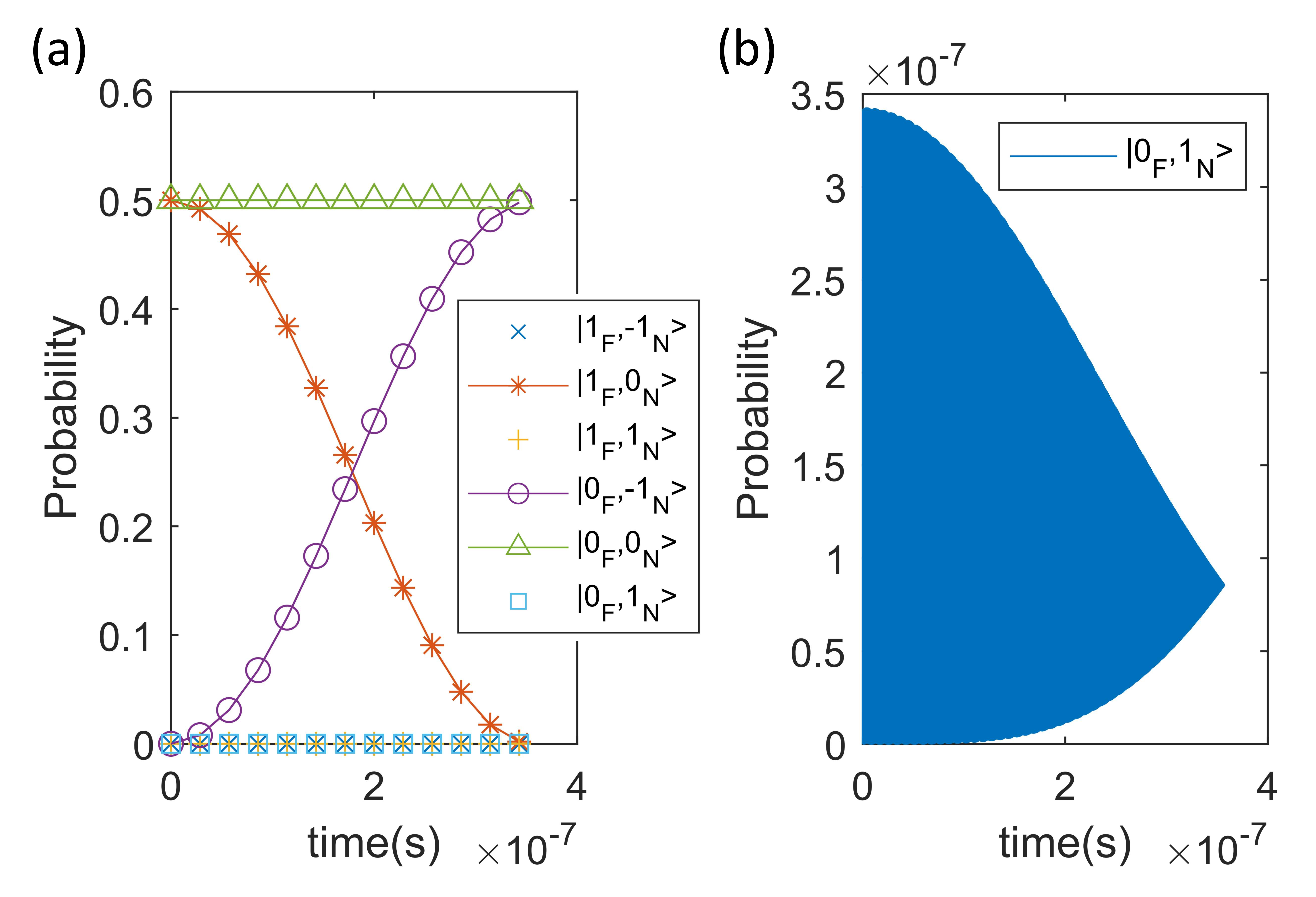} \caption{(a) Dynamics of the probabilities of the basis states of the quantum
memory during a state transfer process for the initial state $\left|\Phi_{\nicefrac{1}{2}}\right\rangle =\sqrt{\nicefrac{1}{2}}\left|1_{F},0_{N}\right\rangle +\sqrt{\nicefrac{1}{2}}\left|0_{F},0_{N}\right\rangle $,
and coupling strength 700 kHz. During the transfer process, $\left|1_{F},0_{N}\right\rangle $
is transferred to $\left|0_{F},-1_{N}\right\rangle $, while $\left|0_{F},0_{N}\right\rangle $
is unchanged. (b) Due to the detuning between the $\left|0_{N}\right\rangle $
to $\left|1_{N}\right\rangle $ transition and the FQ transition frequency,
state $\left|0_{F},1_{N}\right\rangle $ has a negligible probability
$(\sim10^{-7})$ during the process.}
\label{fig:3} 
\end{figure}

Following the exposition of the system parameters, we now continue
to show the numerical results and we plot the dynamics of the transfer
process in Fig.~\ref{fig:3}. 
We choose the case where the effective coupling strength is 700 kHz
and where the initial state is $\left|\Phi_{\nicefrac{1}{2}}\right\rangle =\sqrt{\nicefrac{1}{2}}\left|1_{F},0_{N}\right\rangle +\sqrt{\nicefrac{1}{2}}\left|0_{F},0_{N}\right\rangle $
since this state is, in a more realistic situation considered later,
influenced the most by the dephasing effect. 
During the transfer process (whose duration is 0.36 $\mathrm{\mu}$s),
$\left|1_{F},0_{N}\right\rangle $ is transferred to $\left|0_{F},-1_{N}\right\rangle $,
while $\left|0_{F},0_{N}\right\rangle $ is left unchanged. The population
of the other states remains zero, except that the $\left|0_{F},1_{N}\right\rangle $
state has a small probability $(\sim10^{-7})$ as shown in Fig.~\ref{fig:3}(b).
This small probability is due to the detuning between the transition
frequency from $\left|0_{N}\right\rangle $ to $\left|1_{N}\right\rangle $
and the frequency of the FQ, and one can reduce this probability further
by making the detuning larger.

To simulate the state transfer and storage processes in a more realistic
setting, we use the master equation \cite{Carmichael1999}, which
takes into account both spontaneous decay and pure dephasing of the
FQ and the NV-center spin, 
\begin{align}
\frac{d\rho}{dt}= & -i[H_{\mathrm{eff}},\rho]\nonumber \\
 & +\frac{\gamma_{F}^{(s)}}{2}\mathcal{L}[\sigma_{F}^{(-)}]+\frac{\gamma_{F}^{(p)}}{2}\mathcal{L}[\left|1_{F}\right\rangle \left\langle 1_{F}\right|]\nonumber \\
 & +\sum_{j=\pm1}\frac{\gamma_{N,(j)}^{(s)}}{2}\mathcal{L}[S_{N,(j)}^{(-)}]+\frac{\gamma_{N,(j)}^{(p)}}{2}\mathcal{L}[\left|j_{N}\right\rangle \left\langle j_{N}\right|],\label{eq:master_eq}
\end{align}
where $\mathcal{L}[O]=2O\rho O^{\dagger}-\rho O^{\dagger}O-O^{\dagger}O\rho$,
is the Lindblad superoperator, and $\gamma_{q}^{(s)}$ and $\gamma_{q}^{(p)}$
are the spontaneous decay and the dephasing rates, respectively, of
the species $q$. Note that they are related to the relaxation time
$T_{1}$ and decoherence time $T_{2}$ as $T_{1}=\nicefrac{1}{\gamma^{(s)}}$
and $T_{2}=\nicefrac{2}{\left(\gamma^{(s)}+2\gamma^{(p)}\right)}$,
respectively \cite{Carmichael1999}. It has been reported recently
that both the intrinsic $T_{1}^{*}$ and $T_{2}^{*}$ of FQ's could
be about 10 $\textrm{\ensuremath{\mu}s}$ at 33 mK \cite{Stern2014}
and the value of the intrinsic $T_{2}^{*}$ of an NV-center spin could
be about 90 $\mathrm{\mu s}$ at room temperature \cite{Ishikawa2012}.
Furthermore, it has been shown that by applying dynamical decoupling
pulse sequences, $T_{1}$ of ensembles of NV spins could be more than
10 sec and $T_{2}$ could be about 0.6 sec at 77 K \cite{Bar-Gill2013}.
In the simulations, the relevant decoherence times in the state transfer
stage are the intrinsic $T_{1}^{*}$ and $T_{2}^{*}$ times of the
FQ and the NV-center spin. In the state storage stage, however, the
NV-center spin is effectively decoupled from the YIG and FQ due to
the large detuning and thus dynamical decoupling pulse sequences can
be applied to protect the NV from decoherence to maintain the transferred
state. Consequently, we can use the $T_{1}$ and $T_{2}$ values of
the NV-center spin measured using dynamical decoupling \cite{Bar-Gill2013}
to estimate the fidelity in the state storage stage. Furthermore,
the decoherence times of a NV center solely due to a coupling to a
ferromagnet YIG have been estimated in Ref.~\cite{Trifunovic2013},
and theses times depend sensitively on the ratio of the magnon excitation
gap to the YIG temperature. It has been shown that for a magnon gap
of 100 $\mathrm{\mu}$eV and a temperature of 0.1 K, these times are
typically much larger than the (intrinsic) decoherence times of the
NV \cite{Trifunovic2013}. In other words, the induced decoherence
solely due to coupling to the YIG for temperatures smaller than the
magnon excitation gap is nondetrimental \cite{Trifunovic2013}. In
our scheme, all of the components (the FQ, YIG, and NV) of the qubit
and quantum memory are at the same low temperature as that of the
FQ. Moreover, because of the small size of the YIG and the applied
magnetic field, this ratio of the magnon excitation gap to the temperature
in our scheme is even bigger than that used for estimation in Ref.~\cite{Trifunovic2013}.
As a result, the effect of the induced decoherence solely due to coupling
to the YIG will be neglected in our simulations.

The effect of the linewidth of the YIG nanosphere can also be neglected.
The linewidths of the KM of a single YIG sphere with submillimeter
size have been measured to be about 1\textendash 2 MHz \cite{Tabuchi2014,Zhang2015}.
Although linewidth measurement on a single YIG nanosphere is, to our
knowledge, not available so far, a close case of YIG nanodisks with
thickness about $20$ nm and diameter ranging from $\sim$300\textendash 700
nm has been reported in Ref.~\cite{Hahn2014}. There, the linewidth
of the uniform mode, i.e., the lowest-energy ferromagnetic resonance
mode of a nanodisk, was measured to be about 7 MHz for a nanodisk
with diameter $700$ nm at frequency 8.2 GHz. However, since this
measurement was performed at room temperature and the nanodisks with
different diameters are arranged in a row with 3 $\textrm{\ensuremath{\mu}m}$
spacing (i.e., not completely a single-disk measurement), one may
expect that the linewidth could be narrower if the measurements were
performed for a actual single nanodisk at a low temperature of tens
of mK and at the frequency down to the value of $\sim2$ GHz as in
our proposal. Furthermore, the linewidths of the nanodisks do not
change much with the diameter \cite{Private2018Klein}, at least within
the range investigated in Ref.~\cite{Hahn2014}. One may thus expect
that the linewidth of a YIG nanosphere without surface defects at
low temperature could be similar or at most at a few MHz level, which
is still much smaller than the detuning ($\geq170$ MHz) between the
YIG nanosphere and other quantum systems in our proposal. Consequently,
the effect of the linewidth of the KM of the YIG nanosphere does not
appreciably affect the virtual excitation or virtual coupling picture
in our proposal and thus is neglected in the subsequent calculations
after the degrees of freedoms of the YIG are traced out.

\begin{table*}
\begin{ruledtabular}
\caption{Fidelity at different steps for the initial states of $\left|\Phi_{1}\right\rangle =\left|1_{F},0_{N}\right\rangle $
and $\left|\Phi_{\nicefrac{1}{2}}\right\rangle =\sqrt{\nicefrac{1}{2}}\left|1_{F},0_{N}\right\rangle +\sqrt{\nicefrac{1}{2}}\left|0_{F},0_{N}\right\rangle $
with the storage time 10 ms.}
\begin{tabular}{cccccc}
Initial state  & $g_{F-N}^{\mathrm{(}eff)}$  & $T_{2}^{*}$  & $F(\textrm{FQ}\rightarrow\textrm{NV})$  & $F(\textrm{Storage})$  & $F(\textrm{NV}\rightarrow\textrm{FQ})$\tabularnewline
\hline 
\multirow{4}{*}{$\left|\Phi_{\nicefrac{1}{2}}\right\rangle $} & \multirow{2}{*}{700 kHz} & 90 $\mathrm{\mu s}$ & 0.9689 & 0.9598 & 0.9318\tabularnewline
 &  & 20 $\mathrm{\mu s}$ & 0.9627 & 0.9548 & 0.9218\tabularnewline
\cline{2-6} 
 & \multirow{2}{*}{350 kHz} & 90 $\mathrm{\mu s}$ & 0.9421 & 0.9363 & 0.8880\tabularnewline
 &  & 20 $\mathrm{\mu s}$ & 0.9307 & 0.9270 & 0.8709\tabularnewline
\hline 
\multirow{4}{*}{$\left|\Phi_{1}\right\rangle $} & \multirow{2}{*}{700 kHz} & 90 $\mathrm{\mu s}$ & 0.9317 & 0.9284 & 0.8653\tabularnewline
 &  & 20 $\mathrm{\mu s}$ & 0.9268 & 0.9239 & 0.8562\tabularnewline
\cline{2-6} 
 & \multirow{2}{*}{350 kHz} & 90 $\mathrm{\mu s}$ & 0.8695 & 0.8668 & 0.7537\tabularnewline
 &  & 20 $\mathrm{\mu s}$ & 0.8608 & 0.8581 & 0.7386\tabularnewline
\end{tabular}\label{tab:1} 
\end{ruledtabular}

\end{table*}

\begin{table*}
\begin{ruledtabular}
\caption{Fidelity $F=\sqrt{\left\langle \psi_{t}\right|\rho\left|\psi_{t}\right\rangle }$
after the transfer process with different lengths and types of the
rise times for the initial states of $\left|\Phi_{1}\right\rangle =\left|1_{F},0_{N}\right\rangle $
and $\left|\Phi_{\nicefrac{1}{2}}\right\rangle =\sqrt{\nicefrac{1}{2}}\left|1_{F},0_{N}\right\rangle +\sqrt{\nicefrac{1}{2}}\left|0_{F},0_{N}\right\rangle $.}
\begin{tabular}{cccccc}
State  & $g_{F-N}^{({\rm eff)}}$  & $T_{2}^{*}$  & rise-time func.  & $F(\textrm{4 ns})$  & $F(\textrm{10 ns})$\tabularnewline
\hline 
\multirow{8}{*}{$\left|\Phi_{\nicefrac{1}{2}}\right\rangle $} & \multirow{4}{*}{700 kHz} & \multirow{2}{*}{90 $\mathrm{\mu s}$} & exponential  & 0.9677 & 0.9647\tabularnewline
 &  &  & linear  & 0.9672 & 0.9639\tabularnewline
\cline{3-6} 
 &  & \multirow{2}{*}{20 $\mathrm{\mu s}$} & exponential  & 0.9613 & 0.9581\tabularnewline
 &  &  & linear  & 0.9608 & 0.9573\tabularnewline
\cline{2-6} 
 & \multirow{4}{*}{350 kHz} & \multirow{2}{*}{90 $\mathrm{\mu s}$} & exponential  & 0.9412 & 0.9395\tabularnewline
 &  &  & linear  & 0.9410 & 0.9392\tabularnewline
\cline{3-6} 
 &  & \multirow{2}{*}{20 $\mathrm{\mu s}$} & exponential  & 0.9296 & 0.9278\tabularnewline
 &  &  & linear  & 0.9295 & 0.9275\tabularnewline
\hline 
\multirow{8}{*}{$\left|\Phi_{1}\right\rangle $} & \multirow{4}{*}{700 kHz} & \multirow{2}{*}{90 $\mathrm{\mu s}$} & exponential  & 0.9294 & 0.9241\tabularnewline
 &  &  & linear  & 0.9288 & 0.9228\tabularnewline
\cline{3-6} 
 &  & \multirow{2}{*}{20 $\mathrm{\mu s}$} & exponential  & 0.9246 & 0.9193\tabularnewline
 &  &  & linear  & 0.9240 & 0.9180\tabularnewline
\cline{2-6} 
 & \multirow{4}{*}{350 kHz} & \multirow{2}{*}{90 $\mathrm{\mu s}$} & exponential  & 0.8678 & 0.8648\tabularnewline
 &  &  & linear  & 0.8677 & 0.8646\tabularnewline
\cline{3-6} 
 &  & \multirow{2}{*}{20 $\mathrm{\mu s}$} & exponential  & 0.8591 & 0.8561\tabularnewline
 &  &  & linear  & 0.8590 & 0.8559\tabularnewline
\end{tabular}\label{tab:2} 
\end{ruledtabular}

\end{table*}

We then take the values of the decoherence and relaxation times at
higher temperatures \cite{Ishikawa2012,Bar-Gill2013} to make a conservative
evaluation of the performance of our quantum memory scheme through
the fidelity of the state 
\begin{equation}
F=\sqrt{\left\langle \Psi\right|\rho\left|\Psi\right\rangle },\label{eq:Fidelity}
\end{equation}
where $\left|\Psi\right\rangle $ is the target state and $\rho$
is the actual system density matrix. We can transform the Hamiltonian
to the rotating frame to obtain $H_{\mathrm{int}}$ as in Eqs.~(\ref{eq:H_int})\textendash (\ref{eq:H_int_details}).
Since during the storage stage the system is tuned to be off-resonant,
i.e., $\delta_{B,(-1)}\gg g_{(-1)}$, the total system approximately
undergoes free evolution during this stage. The fidelities of the
quantum state memory for initial states $\left|\Phi_{1}\right\rangle $
and $\left|\Phi_{\nicefrac{1}{2}}\right\rangle $ are shown in Table~\ref{tab:1},
in which results that make use of more conservative values for $T_{2}^{*}=20$
$\mathrm{\mu s}$ for the NV center are also presented. The initial
states $\left|\Phi_{1}\right\rangle $ and $\left|\Phi_{\nicefrac{1}{2}}\right\rangle $
are chosen because they are influenced the most by the spontaneous
decay and dephasing effect, respectively. We have also simulated for
different initial states of $\left|\Phi_{0}\right\rangle =\left|0_{F},0_{N}\right\rangle $
, $\left|\Phi_{\nicefrac{1}{3}}\right\rangle =\sqrt{\nicefrac{2}{3}}\left|1_{F},0_{N}\right\rangle +\sqrt{\nicefrac{1}{3}}\left|0_{F},0_{N}\right\rangle $,
$\left|\Phi_{\nicefrac{1}{4}}\right\rangle =\sqrt{\nicefrac{3}{4}}\left|1_{F},0_{N}\right\rangle +\sqrt{\nicefrac{1}{4}}\left|0_{F},0_{N}\right\rangle $,
and $\left|\Phi_{\nicefrac{1}{5}}\right\rangle =\sqrt{\nicefrac{4}{5}}\left|1_{F},0_{N}\right\rangle +\sqrt{\nicefrac{1}{5}}\left|0_{F},0_{N}\right\rangle $,
and the result shows that $\left|\Phi_{1}\right\rangle $ has the
worst fidelity. This is because during the transfer stage, the main
factor causing infidelity is the decoherence of the FQ, and $T_{1}^{*}$
and $T_{2}^{*}$ of the FQ is about the same in our case so that the
spontaneous decay rate is larger than the dephasing rate. Furthermore,
in the transfer stage, switches take place between $\left|\Phi_{1}\right\rangle =\left|1_{F},0_{N}\right\rangle $
and $\left|0_{F},-1_{N}\right\rangle $, while the state $\left|\Phi_{0}\right\rangle =\left|0_{F},0_{N}\right\rangle $
is unchanged. When the portion of $\left|\Phi_{1}\right\rangle $
in a general initial state decays into $\left|0_{F},0_{N}\right\rangle $,
the state transfer process of that portion will stop and will cause
infidelity. This results in the initial state $\left|\Phi_{1}\right\rangle $
being the worst possible case for the parameters we used. 
Nevertheless, if the effective coupling is stronger through the use
of a YIG moderator containing more spins or if the FQ possesses a
longer coherence time \cite{Yan2016}, 
the fidelity can be appreciably enhanced.

\begin{figure}
\includegraphics[width=1\columnwidth]{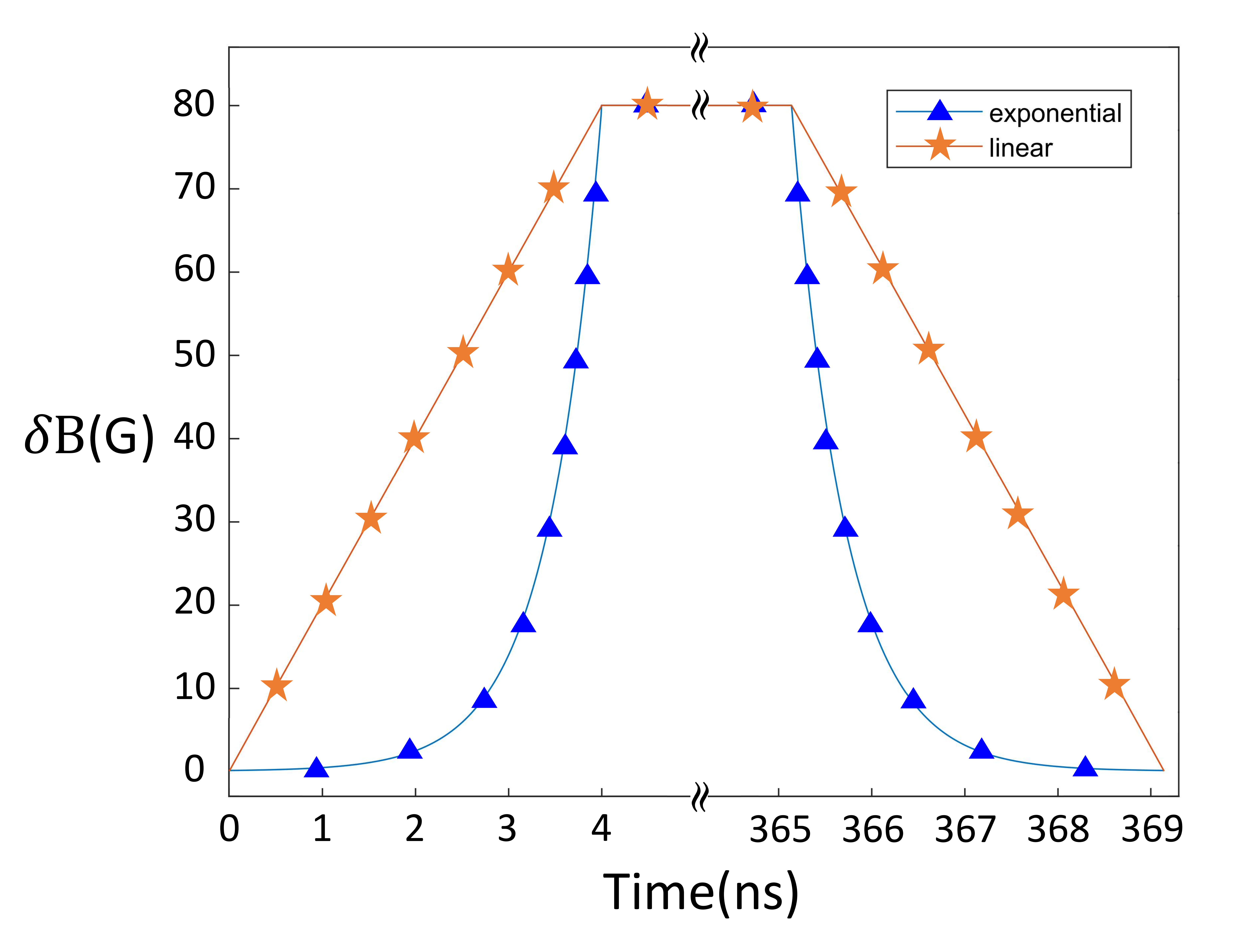} \caption{Temporal variation of the magnetic field with linear or exponential
ramping from off-resonance (0 G) to resonance (80 G) in 4 ns at the
beginning, and using an inverse ramping at the end of the state transfer
stage. Only the rise-time and fall-time regimes are shown and the
storage stage, during which the magnetic field is fixed, is not shown. }
\label{fig:4} 
\end{figure}

In our scheme, the state transfer interaction can be effectively turned
on and off depending on whether or not the NV storage qubit is resonant
with the FQ. We thus can attempt to consider engineering a near-perfect
step function of the external magnetic field $\delta B$ from 0 G
(off) to 80 G (on). We choose the maximum magnetic field strength
such that it is still lower than the critical field of the FQ (the
critical field is 100 G for FQ made of aluminum; could be higher if
made of other superconductor)\cite{Kittel2004}. However, due to technical
limits on the charging and discharging times of circuits, the ramping
of the magnetic field cannot be instantaneous and we take this rise
time into account. We assume a variation of the magnetic field over
200 G in 10 ns, similar to what has been reported in experiments \cite{Salaoru2007}.
In our simulation, $\delta B$ is switched from 0 to 80 G with either
a linear or exponential ramping over a duration of 4 and 10 ns (see
Fig.~\ref{fig:4}). The results shown in Table~\ref{tab:2} indicate
that the linear ramping has slightly lower fidelity than the exponential
ramping. This is due to the fact that the linear ramping makes the
system stay in the near-resonance regime longer and thus subject to
FQ decoherence longer. Shorter rise times of the magnetic field can
also improve the fidelity or, alternatively, one can fine tune the
transfer time to correct the rise-time and fall-time effects.

\section{CONCLUSION}

\label{sec:Conclusion}

We have demonstrated how to couple a superconducting FQ with a single
electron spin of a NV-center spin via a collective KM in a ferromagnetic
material, YIG. This scheme enhances the effective coupling between
the FQ and the NV-center spin, allowing the single-NV-center spin
to interact with the FQ at a longer spatial distance. This provides
greater flexibility for the design of hybrid quantum systems. We have
proposed a protocol for quantum state memory and presented a quantitative
analysis of the state transfer, taking into consideration the possible
decay channels and imperfect technical issues. This YIG architecture
can be used not only as a quantum memory but also as a quantum transducer
that couples a single-NV-center spin with other kinds of qubits or
with a magnetic field. 
\begin{acknowledgments}
G.D.L. and H.S.G. acknowledge support from the the Ministry of Science
and Technology of Taiwan under Grants No.~MOST 105-2112-M-002-015-MY3
and No.~MOST 106-2112-M-002-013-MY3, from the National Taiwan University
under Grant No.~NTU-CCP-106R891703, and from the thematic group program
of the National Center for Theoretical Sciences, Taiwan. J.T. acknowledges
support from the Center of Excellence in Engineered Quantum Systems. 
\end{acknowledgments}

\appendix

\section{Coupling with magnons}

\label{sec:Coupling with magnons}

Here we describe how the magnons in YIG couple to the FQ or the single-NV
spin, and give a detailed derivation of their coupling strengths.
The Hamiltonian of a FQ can be written as $H_{F}=\frac{1}{2}\omega_{F}\sigma_{F}^{(z)}+\frac{1}{2}\epsilon\sigma_{F}^{(x)},$
where $\omega_{F}$ is the energy of the tunnel splitting, $\epsilon=2I_{p}\left(\Phi-\Phi_{0}/2\right)$
is the energy bias, $I_{p}$ is the persistent current of the FQ,
$\Phi_{0}$ is the fluxon, and $\Phi$ is the external flux. If $\Phi$
is tuned to the optimal point with $\Phi=\Phi_{0}/2$ so that $\epsilon=0$,
then one has $H_{F}=\frac{1}{2}\omega_{F}\sigma_{F}^{(z)}$. As discussed
in the main text, the FQ-YIG coupling comes from the Zeeman-like interaction
of the spins in the YIG experienced in the magnetic field $\mathbf{B}_{F}(r)$
produced by the persistent current of the FQ (see Fig.~\ref{fig:1}).
Since the persistent current carried by the side wire of the FQ loop
near the YIG is along the $z$ axis, the magnetic field $\mathbf{B}_{F}(r)=\left(\frac{\mu_{0}}{2\pi r}\right)I_{p}\sigma_{F}^{(x)}$
generated by this persistent current is in the $x$ axis of the YIG
\cite{Douce2015}. Then the coupling Hamiltonian reads 
\begin{align}
H_{FY} & =-\sum_{r}\gamma_{e}\mathbf{B}_{F}(r_{F})\cdot\boldsymbol{S}_{r}\nonumber \\
 & =-\sum_{r}\left(\frac{\mu_{0}\gamma_{e}I_{p}}{2\pi r_{F}}\right)\sigma_{F}^{(x)}S_{r}^{(x)},\label{eq:H_FY}
\end{align}
where $r_{F}$ is the distance between a spin in the YIG and the FQ.
The YIG-NV coupling Hamiltonian through the dipole-dipole interaction
is written as 
\begin{align}
H_{YN}= & -\frac{\mu_{0}}{4\pi}\gamma_{e}^{2}\sum_{r_{N}}\frac{3\left(\boldsymbol{S}_{r}\cdot\hat{\boldsymbol{r}}_{N}\right)\left(\boldsymbol{\sigma}_{N}\cdot\hat{\boldsymbol{r}}_{N}\right)-\left(\boldsymbol{S}_{r}\cdot\boldsymbol{\sigma}_{N}\right)}{2r_{N}^{3}}.\label{eq:H_YN}
\end{align}
Due to the external magnetic field $B$ applied in the $z$ direction,
which makes the frequencies $\omega_{N}$ and $\gamma_{e}B$ much
larger than the dipole-dipole coupling strength, we can apply the
secular approximation to rewrite Eqs.~(\ref{eq:H_YN}) as 
\begin{align}
H_{YN}= & -\frac{\mu_{0}}{8\pi}\gamma_{e}^{2}\hbar\sum_{r}\left(\frac{3\mathrm{cos^{2}}\theta_{r_{N}}-1}{r_{N}^{3}}\right)\left[3S_{r}^{(z)}\sigma_{N}^{(z)}-\boldsymbol{S}_{r}\cdot\boldsymbol{\sigma}_{N}\right],\label{eq:H_YN_simplified}
\end{align}
where $r_{N}$ is the distance between a spin in the YIG and the NV
spin, and $\theta_{r_{N}}$ is the angle between the vector, which
connects the NV and the spin in YIG, and the direction of the external
magnetic field. We can rewrite this Hamiltonian as $H_{YN}\simeq H_{YN}^{\prime}+H_{YN}^{(z)}$,
where 
\begin{align}
H_{YN}^{\prime}= & \sum_{r_{N}}\left(\frac{\mu_{0}\gamma_{e}^{2}\hbar}{4\pi}\right)\left(\frac{3\mathrm{cos^{2}}\theta_{r_{N}}-1}{r_{N}^{3}}\right)\left(S_{r}^{(+)}\sigma_{N}^{(-)}+S_{r}^{(-)}\sigma_{N}^{(+)}\right),\label{eq:H_YN_prime}\\
H_{YN}^{(z)}= & -\sum_{r_{N}}\left(\frac{\mu_{0}\gamma_{e}^{2}\hbar}{4\pi}\right)\left(\frac{3\mathrm{cos^{2}}\theta_{r_{N}}-1}{r_{N}^{3}}\right)S_{r}^{(z)}\sigma_{N}^{(z)}.\label{eq:H_YN_z}
\end{align}

Magnons are low-energy spin-wave excitations, which are used to describe
the collective behavior of the spins in YIG. Since the system of the
quantum memory is at a temperature much lower than the Curie temperature
of YIG, $T_{c}$=559 K, and the YIG is in an off-resonant coupling
regime to the FQ and NV-center spin, one expects the excitation number
to be very small. In this low-temperature and low-excitation regime,
it is convenient to use the Holstein-Primakoff transformation \cite{Holstein1940,Trifunovic2013},
\begin{align}
S_{r}^{(z)}= & -s+a_{r}^{\dagger}a_{r}\approx-s,\label{eq:HPT_z}\\
S_{r}^{(-)}= & \sqrt{2s}\sqrt{1-\frac{n_{r}}{2s}}a_{r}\approx\sqrt{2s}a_{r},\label{eq:HPT_minus}\\
S_{r}^{(+)}= & \left(S_{r}^{(-)}\right)^{\dagger},\label{eq:HPT_plus}
\end{align}
to transform the spin operators in the YIG into bosonic operators,
where each operator is associated with a particle coordinate. Using
the creation and annihilation operators of the magnon modes in the
wave-vector representation,
\begin{align}
a_{k}^{\dagger}= & \frac{1}{\sqrt{N}}\sum_{r}e^{-ik\cdot r}a_{r}^{\dagger},\label{eq:FT_KM_dagger}\\
a_{k}= & \frac{1}{\sqrt{N}}\sum_{r}e^{ik\cdot r}a_{r},\label{eq:FT_KM}
\end{align}
one can then rewrite the Hamiltonian of YIG as Eq.~(\ref{eq:Hs_rewrite}).

Since the coupling strength in $H_{FY}$ is far smaller than $\omega_{F}$
and $\omega_{Y}$, it is valid to use the RWA and Eqs.~(\ref{eq:HPT_z})\textendash (\ref{eq:FT_KM})
to rewrite the FQ-YIG coupling Hamiltonian, given by Eq.~(\ref{eq:H_FY}),
as 
\begin{eqnarray}
H_{FY}^{\prime} & = & -\sum_{r}\left(\frac{\mu_{0}\gamma_{e}I_{p}}{2\pi r_{F}}\right)\left(\sigma_{F}^{(+)}S_{r}^{(-)}+H.C.\right)\nonumber \\
 & \approx & -\sum_{k}\left[g_{FY}(k)\sigma_{F}^{(+)}a_{k}+H.C.\right],\label{eq:H_FY_prime_details}
\end{eqnarray}
with the coupling strength 
\begin{align}
g_{FY}(k)= & \frac{\mu_{0}}{2\pi}\gamma_{e}I_{p}\sqrt{\frac{2s}{N}}\sum_{r_{F}}^{N}\frac{e^{-ik\cdot a}}{r_{F}}.\label{eq:g_FY_details}
\end{align}
Similarly, the YIG-NV coupling reads 
\begin{align}
H_{YN}^{\prime}\simeq & -\sum_{k}\left[g_{YN}(k)a_{k}^{\dagger}\sigma_{N}^{(-)}+H.C.\right]\label{eq:H_YN_prime_details}
\end{align}
where 
\begin{align}
g_{YN}(k)= & -\frac{\mu_{0}}{4\pi}\gamma_{e}^{2}\hbar\sqrt{\frac{2s}{N}}\sum_{r_{N}}^{N}\left(\frac{3\mathrm{cos}^{2}\theta_{r_{N}}-1}{r_{N}^{3}}\right)e^{ik\cdot a},\label{eq:g_YN_details}
\end{align}
is the coupling strength between the YIG and NV-center spin. Using
Eq.~(\ref{eq:HPT_z}), one can also rewrite Eq.~(\ref{eq:H_YN_z})
as 
\begin{align}
H_{YN}^{(z)} & \simeq\delta_{YN}\sigma_{N}^{(z)},\label{eq:H_FY_z_details}
\end{align}
where $\delta_{YN}=\sum_{r_{N}}\frac{\mu_{0}}{4\pi}\gamma_{e}^{2}\hbar\left(\frac{3\mathrm{cos}^{2}\theta_{r_{N}}-1}{r_{N}^{3}}\right)s$
is the induced energy shift to the NV-center storage qubit due to
the coupling with the YIG.

\section{Derivation of the effective Hamiltonian by the Schriffer-Wolff transformation}

\label{sec:Derivation of the effective Hamiltonian by Schriffer-Wolff transformation}
Here we describe here the procedure to derive the effective Hamiltonian
between the the FQ and the NV-center spin. Following Schriffer and
Wolff's approach \cite{Schrieffer1966,Salomaa1988,Bravyi2011}, we
can make a canonical transformation $e^{-\eta}$ on our original Hamiltonian
$H=H_{0}+H_{c}$, where $H_{0}=H_{s}+H_{YN}^{(z)}$ with $H_{s}$
defined in Eq.~(\ref{eq:Hs}), $H_{YN}^{(z)}$ defined in Eq.~(\ref{eq:H_FY_z_details}),
and $H_{c}$ defined in Eq.~(\ref{eq:Hc}). The Hamiltonian after
the transformation reads 
\begin{align}
\tilde{H}= & e^{\eta}He^{-\eta}\nonumber \\
= & H+[\eta,H]+\frac{1}{2!}[\eta,[\eta,H]]+\cdots.\label{eq:CT}
\end{align}
By choosing proper operator $\eta$ satisfying $[H_{0},\eta]=H_{c}$,
one has 
\begin{align}
\tilde{H}\simeq & H_{0}+\frac{1}{2}[\eta,H_{c}].\label{eq:H_eff_details}
\end{align}
In our case,
\begin{align}
\eta= & \lim_{\lambda\rightarrow0}\left[-i\int_{0}^{\infty}H_{c}(t)e^{-\lambda t}dt\right],\label{eq:eta}
\end{align}
and 
\begin{align}
\tilde{H}\simeq & H_{0}-\lim_{\lambda\rightarrow0}\frac{i}{2}\int_{0}^{\infty}[H_{c}(t),H_{c}]e^{-\lambda t}dt,\label{eq:H_eff_eta_details}
\end{align}
where $H_{c}(t)=e^{iH_{0}t}H_{c}e^{-iH_{0}t}$. Since the size of
the YIG is small, the KM of YIG \cite{Tabuchi2015,Zhang2015} is gapped
from the higher-energy modes. Thus, in a low-temperature and virtual-excitation
regime of our scheme, we consider only the KM of the YIG to mediate
the effective coupling strength between the FQ and the NV-center spin.
To obtain the effective Hamiltonian between the FQ and the NV-center
spin without considering the detailed dynamics of the YIG, we trace
out the degrees of freedoms of the YIG , i.e., $H_{\mathrm{eff}}=\langle\tilde{H}\rangle_{Y}$,
with $\left\langle a_{K}^{\dagger}a_{K}\right\rangle _{Y}=n_{K}$
being the mean occupation number of the KM (similar to a mean-field
approximation). This is a good approximation as the YIG is in a low-temperature
and low-excitation regime. Since $H_{c}=H_{FY}^{\prime}+H_{YN}^{\prime}$,
we can rewrite Eq.~(\ref{eq:H_eff_eta_details}) by categorizing
the terms of the commutators after a trace over the YIG degrees of
freedom into two types,
\begin{align}
H_{\mathrm{eff}}\equiv & H_{0}-\left(\delta H_{s}+\delta H_{c}\right)\label{eq:H_eff_eta_details_rewrite}
\end{align}

The first type in Eq.~(\ref{eq:H_eff_eta_details_rewrite}) reads
\begin{align}
\delta H_{s}= & \lim_{\lambda\rightarrow0}\frac{i}{2}\int_{0}^{\infty}\left\langle [H_{FY}^{\prime}(t),H_{FY}^{\prime}]+[H_{YN}^{\prime}(t),H_{YN}^{\prime}]\right\rangle _{Y}e^{-\lambda t}dt\label{U_s}\\
= & \frac{1}{2}\delta_{F}\sigma_{F}^{(z)}+\frac{1}{2}\delta_{N}\sigma_{N}^{(z)},\label{eq:U_s}
\end{align}
where 
\begin{align}
\delta_{F}= & g_{FY}^{2}(\omega_{K})\left(\frac{1}{\omega_{F}-\omega_{K}}\right),\label{eq:delta_F_details}\\
\delta_{N}= & g_{YN}^{2}(\omega_{K})\left[\frac{1}{\left(\omega_{F}+\delta_{YN}\right)-\omega_{K}}\right],\label{eq:delta_N_details}
\end{align}
with $\omega_{K}$ denoting the frequency of the KM, are the energy
shifts of the qubits of the FQ and NV systems, respectively. We demonstrate
how to derive Eq.~(\ref{eq:U_s}) from Eq.~(\ref{U_s}) by calculating
the term containing $\delta_{F}$ explicitly, and then the other term
containing $\delta_{N}$ can be obtained in a similar way. The commutator
in the first term of Eq.~(\ref{U_s}) considering only the KM is 
\begin{widetext}
\begin{align}
\langle[H_{FY}^{\prime}(t),H_{FY}^{\prime}]\rangle_{Y}= & g_{FY}^{2}\left\langle [\sigma_{F}^{(+)}(t)a_{K}(t),\sigma_{F}^{(-)}a_{K}^{\dagger}]+[\sigma_{F}^{(-)}(t)a_{K}^{\dagger}(t),\sigma_{F}^{(+)}a_{K}]\right\rangle _{Y}\nonumber \\
= & g_{FY}^{2}\left\langle e^{i(\omega_{F}-\omega_{K})t}\left[\sigma_{F}^{(-)}\sigma_{F}^{(+)}+\sigma_{F}^{(z)}\left(a_{K}^{\dagger}a_{K}+1\right)\right]\right.\nonumber \\
 & \left.+e^{-i(\omega_{F}-\omega_{K})t}\left[-\sigma_{F}^{(-)}\sigma_{F}^{(+)}-\sigma_{F}^{(z)}\left(a_{K}^{\dagger}a_{K}+1\right)\right]\right\rangle _{Y}\nonumber \\
= & g_{FY}^{2}\left\{ e^{i(\omega_{F}-\omega_{K})t}\left[\sigma_{F}^{(-)}\sigma_{F}^{(+)}+\sigma_{F}^{(z)}\left(n_{K}+1\right)\right]\right.\nonumber \\
 & \left.+e^{-i(\omega_{F}-\omega_{K})t}\left[-\sigma_{F}^{(-)}\sigma_{F}^{(+)}-\sigma_{F}^{(z)}\left(n_{K}+1\right)\right]\right\} .\label{eq:H_FY(t),H_FY}
\end{align}
Then, integrating it over time as in Eq.~(\ref{U_s}), one obtains
\begin{align}
\lim_{\lambda\rightarrow0}\frac{i}{2}\int_{0}^{\infty}\langle[H_{FY}^{\prime}(t),H_{FY}^{\prime}]\rangle_{Y}e^{-\lambda t}dt= & \frac{i}{2}g_{FY}^{2}\lim_{\lambda\rightarrow0}\int_{0}^{\infty}\left\{ e^{i(\omega_{F}-\omega_{K})t}\left[\sigma_{F}^{(-)}\sigma_{F}^{(+)}+\sigma_{F}^{(z)}\left(n_{K}+1\right)\right]\right.\nonumber \\
 & \left.+e^{-i(\omega_{F}-\omega_{K})t}\left[-\sigma_{F}^{(-)}\sigma_{F}^{(+)}-\sigma_{F}^{(z)}\left(n_{K}+1\right)\right]\right\} e^{-\lambda t}dt\nonumber \\
= & g_{FY}^{2}\lim_{\lambda\rightarrow0}\left(\frac{1}{\omega_{F}-\omega_{K}+i\lambda}\right)\left[\frac{1}{2}\left(2n_{K}+1\right)\sigma_{F}^{(z)}+\frac{1}{2}I_{F}\right]\nonumber \\
= & g_{FY}^{2}\left(\frac{1}{\omega_{F}-\omega_{K}}\right)\left[\frac{1}{2}\left(2n_{K}+1\right)\sigma_{F}^{(z)}+\frac{1}{2}I_{F}\right]\nonumber \\
\simeq & \frac{1}{2}g_{FY}^{2}\left(\frac{1}{\omega_{F}-\omega_{K}}\right)\sigma_{F}^{(z)},\label{eq:intH_FY(t),H_FY}
\end{align}
where the constant energy term containing the identity operator $I_{F}$
can be ignored, and $n_{K}\rightarrow0$ since the system is operated
in a virtual-excitation regime. Equation (\ref{eq:intH_FY(t),H_FY})
is the first term of Eq.~(\ref{eq:U_s}). Similarly, the second term
of Eq.~(\ref{U_s}) can be calculated and yields the term containing
$\delta_{N}$ in Eq.~(\ref{eq:U_s}). 

The other type $\delta H_{c}$ in Eq.~(\ref{eq:H_eff_eta_details_rewrite})
represents the effective coupling between the FQ and NV-center spin:
{} 
\begin{align}
\delta H_{c}= & \lim_{\lambda\rightarrow0}\frac{i}{2}\int_{0}^{\infty}\left\langle [H_{FY}^{\prime}(t),H_{YN}^{\prime}]+[H_{YN}^{\prime}(t),H_{FY}^{\prime}]\right\rangle _{Y}e^{-\lambda t}dt\label{eq:dHc}\\
= & g_{FN,\mathrm{eff}}\left(\sigma_{F}^{(+)}\sigma_{N}^{(-)}+H.C.\right),\label{eq:U_c}
\end{align}
with 
\begin{align}
g_{FN,\mathrm{eff}}= \frac{1}{2}g_{FY}(K)g_{YN}(K)
 \left[\frac{1}{\omega_{F}-\omega_{K}}+\frac{1}{\left(\omega_{N}+\delta_{YN}\right)-\omega_{K}}\right].\label{eq:g_FN_eff_details}
\end{align}
We now show how to obtain Eq.~(\ref{eq:U_c}) from Eq.~(\ref{eq:dHc}).
Following the same approach as in Eqs.~(\ref{eq:H_FY(t),H_FY}) and
(\ref{eq:intH_FY(t),H_FY}), the first term of the commutator in Eq.~(\ref{eq:dHc})
reads 

\begin{align}
\lim_{\lambda\rightarrow0}\frac{i}{2}\int_{0}^{\infty}\langle[H_{FY}^{\prime}(t),H_{YN}^{\prime}]\rangle_{Y}e^{-\lambda t}dt= & \lim_{\lambda\rightarrow0}\frac{i}{2}\int_{0}^{\infty}g_{FY}g_{YN}\left\langle [\sigma_{F}^{(+)}(t)a_{K}(t),a_{K}^{\dagger}\sigma_{N}^{(-)}]+[\sigma_{F}^{(-)}(t)a_{K}^{\dagger}(t),a_{K}\sigma_{N}^{(+)}]\right\rangle _{Y}e^{-\lambda t}dt\nonumber \\
= & \frac{i}{2}g_{FY}g_{YN}\lim_{\lambda\rightarrow0}\int_{0}^{\infty}\left[e^{i(\omega_{F}-\omega_{K})t}\left(\sigma_{F}^{(+)}\sigma_{N}^{(-)}\right)+e^{-i(\omega_{F}-\omega_{K})t}\left(-\sigma_{F}^{(-)}\sigma_{N}^{(+)}\right)\right]e^{-\lambda t}dt\nonumber \\
= & g_{FY}g_{YN}\lim_{\lambda\rightarrow0}\left[\frac{1}{2}\left(\frac{1}{\omega_{F}-\omega_{K}+i\lambda}\right)\sigma_{F}^{(+)}\sigma_{N}^{(-)}+\frac{1}{2}\left(\frac{1}{\omega_{F}-\omega_{K}-i\lambda}\right)\sigma_{F}^{(-)}\sigma_{N}^{(+)}\right]\nonumber \\
= & \frac{1}{2}g_{FY}g_{YN}\left(\frac{1}{\omega_{F}-\omega_{K}}\right)\left(\sigma_{F}^{(+)}\sigma_{N}^{(-)}+\sigma_{F}^{(-)}\sigma_{N}^{(+)}\right).\label{eq:intH_FY(t),H_YN}
\end{align}

The second term of the commutator in Eq.~(\ref{eq:dHc}) can be evaluated
in a similar way and combining with Eq.~(\ref{eq:intH_FY(t),H_YN})
give the results of Eqs.~(\ref{eq:U_c}) and (\ref{eq:g_FN_eff_details}).
Combining all these results, one arrives at the effective Hamiltonian
of Eq.~(\ref{eq:H_eff}).
\end{widetext}
 \bibliographystyle{apsrev4-1}
\bibliography{Qmemory_Paper1}

\end{document}